\documentclass[twocolumn,showpacs,preprintnumbers]{revtex4}
\usepackage{dcolumn}
\usepackage{bm}
\input epsf

\begin{document}


\title{Dark energy, matter creation and curvature}

\author{V\'{\i}ctor H. C\'{a}rdenas}
\email{victor@dfa.uv.cl}

\affiliation{Departamento de F\'{\i}sica y Astronom\'{i}a, Facultad
de Ciencias, Universidad de Valpara\'{i}so, Av. Gran Bretana 1111,
Valpara\'{i}so, Chile}

\begin{abstract}
The most studied way to explain the current accelerated expansion of
the universe is to assume the existence of dark energy; a new
component that fill the universe, does not clumps, currently
dominates the evolution, and has a negative pressure. In this work I
study an alternative model proposed by Lima et al. \cite{lima96},
which does not need an exotic equation of state, but assumes instead
the existence of gravitational particle creation. Because this model
fits the supernova observations as well as the $\Lambda$CDM model, I
perform in this work a thorough study of this model considering an
explicit spatial curvature. I found that in this scenario we can
alleviate the cosmic coincidence problem, basically showing that
these two components, dark matter and dark energy, are of the same
nature, but they act at different scales. I also shown the
inadequacy of some particle creation models, and also I study a
previously propose new model that overcome these difficulties.
\end{abstract}


\maketitle

\section{Introduction}

Currently the observational evidence coming from supernovae studies
\cite{SNIa}, cosmic background radiation fluctuations \cite{cmbr},
and baryon acoustic oscillations \cite{bao}, set a strong case for a
cosmological (concordance) universe model composed by nearly $70$
percent of a mysterious component called dark energy, responsible
for the current accelerated expansion, nearly $25$ percent of dark
matter, which populates the galaxy halos and a small percentage
(around $4$) is composed by baryonic matter. The nature of these two
dark component remains so far obscure \cite{DEDM}. In the case of
dark matter, we have a set of candidates to be probed with
observations and detections in particle accelerators, however the
case of dark energy is more elusive. This component not only has to
fill the universe at the largest scale homogeneously, and also has
the appropriate order of magnitude to be comparable to dark matter,
but also has to have a negative pressure equation of state,
something that was assumed first to explain inflation in the early
universe, inspired by high energy theories of particle physics, but
which seems to be awkward to appeal for, at these low energy scales.
The alternative way to account for the cosmic acceleration, in the
framework of the standard model, is to consider a modification of
general relativity at large scales \cite{alterGR}.

Some years ago, Prigogine and co-workers \cite{prigogine} presented
a very interesting cosmological model where matter creation takes
place without spoiling adiabatic expansion. This is possible by
adding a work term due to the change of the particle number density
$n$. This work actually suggested for the first time, a way to
incorporate the particle creation process in the context of
cosmology in a self consistent way. In fact, the original claim made
by Zeldovich \cite{zeldov70}, that gravitational particle production
can be described phenomenologically by a negative pressure, is
realize here in a beautiful way. A covariant formulation of the
model was presented for the first time in \cite{CLW90}.

Originally, these considerations were applied in the context of; the
steady-state cosmological model, the warm inflationary scenario and
within the standard inflationary scenario, during the reheating
phase. Based on this work, the studies of cosmological models with
matter creation \cite{lima96} were initiated, and rapidly recognized
to be potentially important to explain dark energy \cite{harko}. In
particular, in Ref.\cite{zsbp01} the authors established a model
where dark energy can be mimicked by self-interactions of the dark
matter substratum. Within the same framework, models of interacting
dark energy and dark matter were proposed \cite{interact}. Actually,
in these studies is possible to have consistently, a universe where
matter creation proceeds within an adiabatic evolution. More
recently, Lima et al.,\cite{lima08} have presented a study of a flat
cosmological model where a transition from decelerated to
accelerated phase exist. They explicitly show that previous models
considered \cite{lima99}, does not exhibit the transition, and study
the observational constraints on the model parameters. Also in
\cite{Steigman:2008bc} the authors modified slightly the model,
adding explicitly a baryonic contribution, which enable them to have
a transition from decelerated to accelerated expansion always.

In this work I consider the non flat extension of this matter
creation cosmological model. This is important because, even in the
$\Lambda$CDM concordance model, there is no clear evidence that
$\Omega_{k}$ is zero, due to the well known degeneracy between
$w(z)$ and $\Omega_{k}$ \cite{curvature}. Furthermore, any
alternative model to dark energy then must consider a non flat
assumption as a prior. This model enable us to explain the current
acceleration of the universe expansion through a fictitious pressure
component coming from changes in the dark matter particle number,
without any exotic contribution. As a bonus, this model enable us to
explain easily the cosmic coincidence problem; it is not strange to
have a similar contribution from these two (commonly differently
regarded) components, because they are just two aspects of one and
the same component; the dark matter.

In the next section I derive the equations of motion in the case of
matter creation. Then, in section III I study the case for non flat
models  and the observational consequences in the models already
knew. After this, I discuss a new particle creation model that
resembles the $\Lambda$CDM model case, showing the transition from a
decelerated to an accelerated expansion without enter in conflict
with the behavior for large z.

\section{The effective negative pressure}

Assuming that the particle number is not conserved
$N^{\mu}_{;\mu}\neq 0$, it leads to a modification of the energy
conservation equation. In fact, assuming $N^{\mu}_{;\mu}=n\Gamma$
where $\Gamma$ is the particle creation rate, and using the FRW
metric we obtain the generalization of the energy conservation
equation,
\begin{equation}\label{firstlaw}
d(\rho V)+ pdV -(h/n)d(nV)=0,
\end{equation}
where $h=(\rho+p)$ is the enthalpy (per unit volume), $n$ is the
number density and $\rho$ is the energy density, and a new equation
\begin{equation}\label{ene}
  \dot{n} + 3 H n = n\Gamma,
\end{equation}
for the non conserved number density. The important thing to stress
here is that the universe evolution continues to be adiabatic, in
the sense that the entropy per particle remains unchanged
($\dot{s}=0$). The extra contribution can be interpreted in
(\ref{firstlaw}) as a non thermal pressure defined as
\begin{equation}\label{pc}
p_c=-\left(\frac{\rho + p}{3H}\right) \Gamma.
\end{equation}
This is the source that produces the acceleration of the
universe expansion. Once the particle number increases with the
volume, we obtain a negative pressure.

Now, because we are considering matter creation, we have to impose
the second law constraint
\begin{equation}\label{second}
dS=\frac{s}{n}d(nV) \geq 0,
\end{equation}
where $s=S/V$ is the entropy density. From (\ref{firstlaw}) we find
that the new set of Einstein equations are
\begin{equation}\label{friedman}
  H^2 + \frac{k}{R^2} =  \frac{8\pi G}{3} \rho
\end{equation}
\begin{equation}\label{rho}
  \dot{\rho} = \frac{\dot{n}}{n}(\rho + p)
\end{equation}
and the previously derived relation (\ref{ene}). The set of
equations (\ref{ene}), (\ref{friedman}) and (\ref{rho}) completely
specified the system evolution. It is also useful to combine
(\ref{pc}) with (\ref{friedman}) and (\ref{rho}) to eliminate $\rho$
and obtain
\begin{equation}\label{eq5}
2\frac{\ddot{R}}{R}+\frac{\dot{R}^2}{R^2}+\frac{k}{R^2}=-8 \pi G(p +
p_c).
\end{equation}
The standard adiabatic evolution is easily recovered: setting
$\Gamma = 0$ implies that $\dot{n}/n=-3H$, which leads to the usual
conservation equation from (\ref{rho}). A class of de Sitter
solution is obtained with $\dot{n}=\dot{\rho}=0$ and arbitrary
pressure $p$. Moreover, there exist another class of solutions where
Eq.(\ref{rho}) enable us to determine the pressure; for example if
$\rho=m n$, with $m$ constant, Eq.(\ref{rho}) implies $p=0$, and
furthermore if $\rho=aT^4$ and $n=bT^3$ implies $p=\rho/3$.

I have to stress here that the main result derived in this section
means that we have a single contribution, which satisfy the
non-relativistic matter equation of state, that describe both dark
matter and dark energy simultaneously, and in this way solves
automatically the coincidence problem. This dark unification
mechanism does not have the problem studied in \cite{STZW}, because
in the cases studied in that paper, for example the Chaplygin gas
model \cite{chaplygin}, the sound velocity of the dark matter is not
zero, leading to instabilities. That happens because the Chaplygin
gas interpolates between the equation of states for dark matter and
dark energy. This does not happens here, because there is just one
equation of state; that of non-relativistic matter.

\section{Matter creation in a non flat universe}

In this section we study a number of models where matter creation
produces cosmic acceleration, in a curved background. We also test
the models using the most recent Supernovae data, the so called
Union 2 set \cite{Union2}. In order to do that test we consider the
comoving distance from the observer to redshift $z$ is given by
\begin{equation}\label{comdistance}
r(z) = \frac{1}{\sqrt{-\Omega_k}}\sin \sqrt{-\Omega_k} \int_0^z
\frac{dz'}{E(z')},
\end{equation}
where $E(z)\equiv H(z)/H_0$. The SNIa data give the distance modulus
$\mu(z)$ which is related to the luminosity distance
$d_L(z)=(1+z)r(z)$, through $\mu(z)\equiv
5\log_{10}[d_L(z)/\texttt{Mpc}]+25$. We fit the SNIa with the
cosmological model by minimizing the $\chi^2$ value defined by
\begin{equation}
\chi_{SNIa}^2=\sum_{i=1}^{557}\frac{[\mu(z_i)-\mu_{obs}(z_i)]^2}{\sigma_{\mu
i}^2},
\end{equation}
where $\mu_{obs}$ is the corresponding observed one.

\subsection{$\beta$ Model}

Let us assume first the particle creation rate evolving as
\begin{equation}\label{asump1}
\Gamma=3H\beta,
\end{equation}
where $\beta$ is a constant. From (\ref{ene}) we finds the solution
for the density number
\begin{equation}\label{sol1}
n(R)= n_0 \left( \frac{R_0}{R}\right)^{3(1-\beta)},
\end{equation}
where $n_0$ is a constant of integration. Assuming $\rho=m n$ as for
non-relativistic matter (that leads to que usual equation of state
$p=0$ as we discussed that at the end of the last section), we get
for the Hubble function
\begin{equation}\label{rhosol}
E^2(z) =  \Omega_m (1+z)^{3(1-\beta)}+(1-\Omega_m)(1+z)^2.
\end{equation}
Clearly, this model does not describe properly a transition from
decelerated to an accelerated expansion phase. In fact, from
(\ref{friedman}) and (\ref{rho}) we obtain:
\begin{equation}\label{addot}
\frac{\ddot{R}}{R}=4 \pi G \rho \left(\beta - \frac{1}{3}\right).
\end{equation}
Depending on the value assumed for $\beta$ we obtain a model that
accelerate forever ($\beta >1/3$) or decelerate forever
($\beta<1/3$).
\begin{figure}[h!]
\centering \leavevmode\epsfysize=6cm \epsfbox{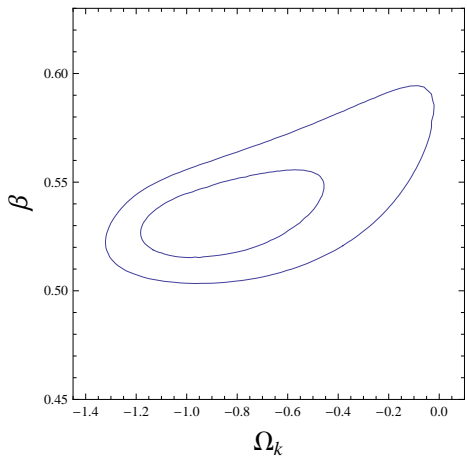}\\
\caption{\label{fig02} Using the Union 2 data set \cite{Union2} we
plot the confidence contours, at $68.27\% $ and $95.45\%$ for the
two parameters of the model: $\beta$ and $\Omega_k$.}
\end{figure}
A Bayesian analysis using the Union 2 data set of SNIa \cite{Union2}
enable us to plot confidence contours for the two parameters of the
model: $\beta$ and $\Omega_k$. This is shown in Fig.\ref{fig02}.
Clearly, the data suggest a large negative value for the curvature
parameter, indicating that in order to fit the SNIa data, the
function $H(z)$ has to have a very distorted form. In any case,
because this model does not present a transition from a decelerated
to an accelerated phase, it can be ruled out immediately.

\subsection{$\gamma$ model}

As is well known from observations of supernovae, a transition from
a decelerated to an accelerated expansion occurs in the recent
history of the universe. Depending on what is used to model dark
energy, different redshift have been obtained between $0.5$ to $1$.
In the context of a model of adiabatic matter creation, this topic
was recently discussed in \cite{lima08} where the authors
specialized in a flat cosmology, where a explicit transition can be
achieved from a decelerated to an accelerated expansion using the
following model
\begin{equation}\label{rate12}
\Gamma = 3 \gamma H_0.
\end{equation}
In this section we generalize this work to non flat universes. In
what follows, a non relativistic matter equation of state is assumed
$\rho=mn$ ($p=0$).

Introducing (\ref{rate12}) in (\ref{ene}) leads to the following
solution

\begin{equation}\label{nder}
n(R)=n_0\left(\frac{R_0}{R}\right)^3 e^{3H_0\gamma (t-t_0)}.
\end{equation}
Is evident the meaning of the subscript zero. Because $\rho=mn$ for
non-relativistic matter and using (\ref{eq5}) we obtain

\begin{equation}\label{ddr2}
\frac{\ddot{R}}{R}=4 \pi G\left(\gamma \frac{H_0}{H}-\frac{1}{3}
\right),
\end{equation}
that clearly allow us to describe an acceleration/deceleration
transition. Writing the right hand side in terms of $ H $ and using
the Hubble equation once more, we can write

\begin{equation}\label{heqt}
\dot{H}+H^2 = \frac{3}{2}\left( H^2+\frac{k}{R^2}\right)\left(\gamma
\frac{H_0}{H}-\frac{1}{3} \right).
\end{equation}
For $k=0$ this equation coincides with that in \cite{lima08}, for
which an explicit form of $H(z)$ was found. In the general case, $ k
\neq 0$, it is not possible to integrate (\ref{heqt}) to obtain a
closed analytic form. Instead, we use $\dot{R}/R=H$ to change the
time derivatives for derivatives of the scale factor $R$, and using
that $R_0/R=1+z$, change again the derivatives in terms of the
redshift. Doing that, we obtain

\begin{equation}\label{edez}
(1+z)E\frac{dE}{dz}-E^2 = \left[\Omega_k (1+z)^2-E^2
\right]\left(\frac{3\gamma}{2E}-\frac{1}{2} \right),
\end{equation}
where $E(z)=H(z)/H_0$. Given the set of parameters $\Omega_k$ and
$\gamma$, we integrate numerically (\ref{edez}) to obtain the
function $E(z)$.
\begin{figure}[h!]
\centering \leavevmode\epsfysize=6cm \epsfbox{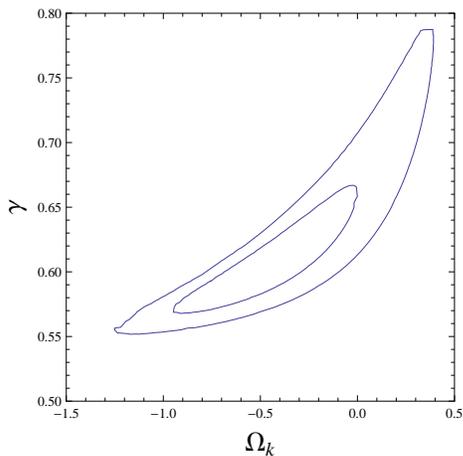}\\
\caption{\label{fig03} Using the Union 2 data set \cite{Union2} we
plot the confidence contours, at $68.27\% $ and $95.45\%$ for the
two parameters of the model: $\gamma$ and $\Omega_k$, by integrating
Eq. (\ref{edez}).}
\end{figure}
A Bayesian analysis using the Union 2 data set of SNIa \cite{Union2}
enable us to plot confidence contours for the two parameters of the
model: $\gamma$ and $\Omega_k$. This is shown in Fig.\ref{fig03}.
Again, as in the previous subsection, the SNIa data suggest a very
large negative value for $\Omega_k$.

\subsection{Problems with these models}

\begin{figure}[h!]
\centering \leavevmode\epsfysize=5cm \epsfbox{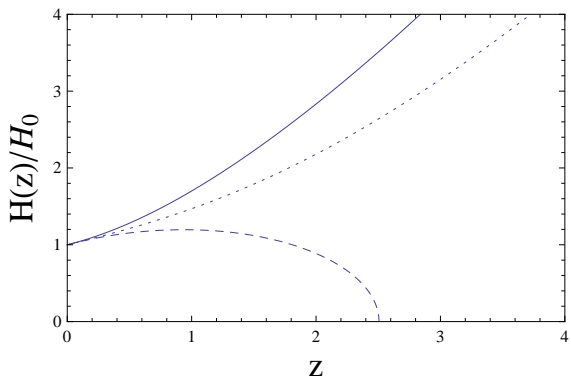}\\
\caption{\label{hzf} The Hubble function in terms of the redshift
for the $\beta$ model (dashed line), the $\gamma$ model (dotted
line) and the flat LCDM model,
$(\Omega_m,\Omega_{\Lambda})=(0.27,0.73)$ (continuos line). }
\end{figure}

A simple way to understand the large negative values obtained for
$\Omega_k$ in the fitting process, which is suggested by the SNIa
data in these models, is by plotting the $H(z)$ function in the
range $z\in (0,4)$. In Fig. \ref{hzf} we display such a plot.
All the three models have two free parameters to be fixed with the
SNIa data. Although at low redshift ($z<0.2$) the three curves
coincide, their evolution with increasing $z$ differs appreciably.
Clearly the $\beta$ model behaves erroneously. Instead, the $\gamma$
model, although it has the correct tendency, it is easily
differentiated from the LCDM at large redshift ($z>0.2$). Because we
are using the SNIa data set, the best fit values for the free
parameters in these models (shown in Fig.\ref{hzf}) adjust
principally the low-$z$ zone, because from the $557$ SNIa data only
$19$ have $z>1$. This is also clear once we try to perform a joint
Bayesian analysis using SNIa plus the BAO and CMB constraints (see
Table I). Using the $\beta$ model, the procedure even does not find
any appropriate configuration. Using the $\gamma$ model, it is
possible to get a good fit ($\chi_{red}=0.994$) using only SNIa+BAO,
however once we add the CMB constraint, it worsens the fit
($\chi_{red}=1.32$).

\begin{table}[h!]
\caption{\label{tab:table0} The $\chi^2$ for the best fit values in
models discussed in this section plus the curved $\Lambda$CDM using
the Union2 data set in the case of a non flat universe model.}
\begin{ruledtabular}
\begin{tabular}{cccc}
Model & SN & SN+BAO & SN+BAO+CMB   \\
\hline
$\beta$ Model & 542.17 & 554.99 & --- \\
$\gamma$ Model & 542.29 & 552.66 & 731.12 \\
$\Lambda$CDMk & 542.55 & 542.65 & 542.65 \\
\end{tabular}
\end{ruledtabular}
\end{table}

\begin{figure}[h!]
\centering \leavevmode\epsfysize=5cm \epsfbox{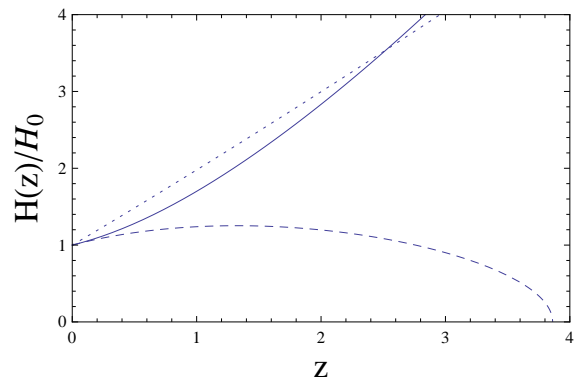}\\
\caption{\label{hzf2} The same as Fig. \ref{hzf}, but this time
considering baryonic dark matter with $\Omega_B=0.042$. See the text
for a discussion. }
\end{figure}

It is therefore clear why in \cite{Steigman:2008bc} the authors add
an explicit contribution for baryons. Adding baryonic dark matter
with $\Omega_B=0.042$ in the previous models, enable us to obtain
better best fit parameters for the models that, for instance
modifying Fig. \ref{hzf} into Fig. \ref{hzf2}.

For example, in the case of the $\gamma$ model, once we add baryons
the best fit parameters improve respect to the case without baryons;
using only SNIa+BAO we get a $\chi_{red}=0.989$, however once we add
the CMB constraint, it worsens the fit again.

\section{Search for a New model}

The previous models have been discussed in several papers, deriving
their consequences in cosmological evolution and also testing their
performance to fit the supernova data. The problem with the $\beta$
model is that it can not describe properly a transition from a
decelerated to an accelerated phase. In contrast, the $\gamma$ model
can do this, however they can only fit well low-redshift
observational constraint (SNIa data with $z<0.2$). As we mentioned
in the last section, if we attempt to include the BAO and CMB
constraint, the whole fit breaks down.

The objective merit of this type of model is that we can mimic a
cosmological constant without use of an exotic component, we just
need to assume the existence of  matter creation that takes place
with a certain rate. In principle, we do not have any fundamental
reason to use a specific form of the matter creation rate. The only
fundamental feature we have to respect is the second law of
thermodynamics (\ref{second}).

In \cite{Cardenas:2008zv} we propose a model that was also discussed
in \cite{Lima:2009ic}. Let us consider this model here as an
example. This model is characterized by the following matter
creation rate
\begin{equation}\label{rate1}
n \Gamma = 3 \alpha H,
\end{equation}
which for $\alpha>0$ satisfy the requirement, because $H(z)$ is a
positive definite function. Inserting in (\ref{ene}) we obtain
\begin{equation}\label{solene}
n(R) = (n_0-\alpha) \frac{R_0^3}{R^3} + \alpha.
\end{equation}
that resembles the combined contribution of a cosmological constant
and dust. The level of fine tuning here to obtain an accelerated
expansion is relatively smaller than in the usual $\Lambda$CDM
model, because in this case, both terms come from the same function,
and we know that we can obtain an accelerated expansion phase after
some period of time from (\ref{friedman}). In fact, from (\ref{pc})
we get $p_c=-m\alpha$, so replacing in (\ref{eq5}) we find
\begin{equation}\label{accelerated}
\frac{\ddot{R}}{R}= \frac{4\pi G m}{3} \left[2\alpha - (n_0 -\alpha)
\left( \frac{R_0}{R}\right)^3 \right].
\end{equation}
Note however that the equation of state has not been modified,
$\rho$ is still the non-relativistic contribution. We do not have to
introduce any exotic component - with a negative pressure - to
describe the current expansion acceleration. Given the current
status of the dark matter and dark energy problem \cite{DEDM}, we
can use this idea as a possible way to understand it. Replacing in
the Hubble equation we obtain
\begin{equation}\label{edez3}
E(z)^2 = \Omega_{\alpha} + (1-\Omega_{\alpha}-\Omega_{k})(1+z)^3 +
\Omega_{k}(1+z)^2,
\end{equation}
where $\Omega_{\alpha}= m\alpha/\rho_{cr}$ and
$\rho_{cr}=3H_0^2/8\pi G$, which is effectively indistinguishable
from a curved LCDM model. Because
$\Omega_{\alpha}=\Omega_{\Lambda}=0.7$, is clear that the
observations implies a positive $\alpha$ for this new model.


In this work I have studied a model which considering small changes
in the total number of particles in our universe, offer a possible
way to understand the current accelerated expansion measurements,
without using any exotic energy component. Also, this scenario
alleviate the cosmic coincidence problem, basically showing that
these two components, dark matter and dark energy, are of the same
nature, but their act at different scales. This way of understand
the SNIa observations, implies that cosmology have a new window to
explore the universe considering matter creation, encoded in the
function $dN/dV$. I have discussed the two models already known in
the literature, extending the analysis to nonflat universes,
performing a bayesian analysis using both SNIa and also BAO and CMB.
I have found that, both the $\beta$ and $\gamma$ models fail to fit
the observations, even considering $\Omega_k \neq 0$. Although
adding an explicit baryonic dark matter contribution may ameliorate
the fit using SNIa+BAO in the $\gamma$ model, adding the CMB
constraint worsens the situation. The statistical analysis was
performed even in the case an explicit form for $H(z)$ was lacking.
I have also studied a previously proposed new model, through a
specific form of the matter creation rate, which is almost
indistinguishable from a LCDM model, which enable us to fit all the
observations, SNIa+BAO+CMB. Although similar to LCDM, the model can
be distinguished through the structure formation process, which will
be studied elsewhere.

\section*{Acknowledgments}

The author want to thank S. del Campo and R. Herrera for useful
discussions, and acknowledges financial support through DIUV project
No. 13/2009, and FONDECYT 1110230.


\end{document}